\documentstyle[fullpage]{article}
\title{\bf Abstract State Machines 1988-1998: Commented ASM Bibliography
}

\author{{\bf Egon B\"{o}rger} and {\bf James K. Huggins} }
\begin{document}
\maketitle

\begin{abstract} 
This is the current version of an annotated bibliography of papers
which deal with or use ASMs. It is compiled to a great extent from
references and annotations provided by the authors of the listed
papers and extends the annotated bibliography which previously appeared
in \cite{Boerger95a}. Comments, additions and corrections are
welcome and should be sent to \texttt{boerger@di.unipi.it} and 
\texttt{huggins@acm.org}\footnote{To appear in {\em Formal Specification
Column} (Ed. H.Ehrig), Bulletin of the EATCS 64, February 1998.}.

\end{abstract}

    Hartmut Ehrig asked the first author to write for this column what
    are the distinguishing features of the ASM approach to specification
    and verification of complex computing systems. In \cite{Boerger95b} an
    attempt had already been made to answer that question by discussing,
    in general comparative terms, some specific features which are
    characteristic for the ASM approach with respect to other well known
    approaches in the literature. That explanation seems to have been
    understood, as shown by the many positive reactions, but even more the
    numerous critical reactions of colleagues in the field who
    felt---rightly---that ASMs put justified doubt on cherished
    denotational, declarative, logical, functional and similar widespread
    beliefs in {\em pure}, i.e. not operational methods. Nevertheless some
    dissatisfaction remained with that paper because the discussion, in a
    sense unavoidably, remained in general terms which have been used
    during the last two or three decades again and again for the
    justification of many other methods.

    The attempt to answer the question in a more {\em concrete} way led
    the two authors of this commented bibliography to systematically
    review again, revising and updating \cite{Boerger95a}, what are
    the achievements and failures of ASM research since the
    discovery of the notion by Yuri Gurevich in 1988. What follows here is
    a way of answering Hartmut Ehrig's question; namely, we try to let the
    research results speak for the method. 

    If somebody really wants to know whether there is anything useful in
    the notion of ASM which has not been covered by competing methods in
    the literature, he or she should try out the method on a challenging
    (not a toy) specification or verification problem. We have no doubt
    that then it will become clear why so much successful research could
    be done in such a short period by a relatively small number of
    researchers, as documented in the commented bibliography below.

Current updates of this bibliography (as well as some of the papers
listed below) will be available on the ASM web sites
\texttt{http://www.eecs.umich.edu/gasm} and
\texttt{http://www.uni-paderborn.de/cs/asm.html}.

\nocite{*}

\bibliographystyle{plain}
\bibliography{annbib}

\begin{thebibliography}{100}

\bibitem{Ahrendt95}
W.~Ahrendt.
\newblock {Von Prolog zur WAM. Verifikation der Prozedur\"ubersetzung mit KIV.}
\newblock Master's thesis, Universit{\"at} Karlsruhe, Karlsruhe, Germany, 1995.
\newblock \medskip \newline In German, starting point for \cite{SchAhr97}. See
  comment to \cite{BoeRos95}.

\bibitem{AnKuPi97}
M.~Anlauff, P.~Kutter, and A.~Pierantonio.
\newblock {Formal Aspects of and Development Environments for Montages}.
\newblock In M.~Sellink, editor, {\em {2nd International Workshop on the Theory
  and Practice of Algebraic Specifications}}, Workshops in Computing,
  Amsterdam, 1997. Springer.
\newblock \medskip \newline A description of the use of Montages
  \cite{KutPie97a} and the {\sc Gem-Mex} tool, with some small examples.

\bibitem{Araujo97}
L.~Araujo.
\newblock {Correctness proof of a Distributed Implementation of Prolog by means
  of Abstract State Machines}.
\newblock {\em {Journal of Universal Computer Science}}, 3(5):416--422, 1997.
\newblock \medskip \newline Building upon \cite{BoeRos95}, a specification and
  a proof of correctness for the Prolog Distributed Processor (PDP), a WAM
  extension for parallel execution of Prolog on distributed memory are
  provided. A preliminary version appeared in 1996 under the title {\em
  Correctness proof of a Parallel Implementation of Prolog by means of Evolving
  Algebras} as Technical Report DIA 21-96 of Dpto. Inform\'atica y
  Autom\'atica, Universidad Complutense de Madrid.

\bibitem{BeaSli97b}
D.~B\`eauquier and A.~Slissenko.
\newblock {On Semantics of Algorithms with Continuous Time}.
\newblock Technical Report 97-15, Dept. of Informatics, Universit\'e Paris-12,
  October 1997.
\newblock \medskip \newline A continuation of \cite{BeaSli97a}. The authors
  consider a class of algorithms with explicit continuous time (a modified
  version of ASMs), a logic which suffices to write requirements specifications
  close to natural language, and the corresponding verification problem, all in
  a single logic. An enhanced logic from that used in \cite{BeaSli97a} is
  presented and used to give a proof of correctness of the Railroad Crossing
  problem \cite{GurHug96}.

\bibitem{BeaSli97a}
D.~B\`eauquier and A.~Slissenko.
\newblock {The Railroad Crossing Problem: Towards Semantics of Timed Algorithms
  and their Model-Checking in High-Level Languages}.
\newblock In M.~Bidoit and M.~Dauchet, editors, {\em TAPSOFT'97: Theory and
  Practice of Software Development, 7th International Joint Conference
  CAAP/FASE}, volume 1214 of {\em LNCS}, pages 201--212. Springer, 1997.
\newblock \medskip \newline The ASM specification of the railroad crossing
  problem \cite{GurHug96} is analyzed to create an appropriate timed-transition
  system, suitable for algorithmic model checking. An early version appeared in
  1995 as Technical Report 96-10 of Dept. of Informatics, Universit\'e
  Paris-12. For a continuation see \cite{BeaSli97b}.

\bibitem{BecPos96}
B.~Beckert and J.~Posegga.
\newblock {leanEA: A Lean Evolving Algebra Compiler}.
\newblock In H.~Kleine B{\"u}ning, editor, {\em {\em Proceedings of the Annual
  Conference of the European Association for Computer Science Logic
  (CSL'95)}\/}, volume 1092 of {\em LNCS}, pages 64--85. Springer, 1996.
\newblock \medskip \newline A 9-line Prolog interpreter for sequential ASMs,
  including discussion of extensions for layered ASMs. A preliminary version
  appeared in April 1995 under the title {\em leanEA: A poor man's evolving
  algebra compiler} as internal report 25/95 of Fakult{\"a}t f{\"u}r
  Informatik, Universit{\"a}t Karlsruhe.

\bibitem{Beierle94}
C.~Beierle.
\newblock {Formal Design of an Abstract Machine for Constraint Logic
  Programming}.
\newblock In B.~Pehrson and I.~Simon, editors, {\em IFIP 13th World Computer
  Congress}, volume I: Technology/Foundations, pages 377--382, Elsevier,
  Amsterdam, the Netherlands, 1994.
\newblock \medskip \newline Proposes a general implementation scheme for CLP(X)
  over an unspecified constraint domain X by designing a generic extension
  WAM(X) of the Warren Abstract Machine and a corresponding generic compliation
  scheme of CLP(X) programs to WAM(X) code. The scheme is based on the
  specification and correctness proof for compilation of Prolog programs in
  \cite{BoeRos95}.

\bibitem{BeiBoe92}
C.~Beierle and E.~B{\"o}rger.
\newblock {Correctness Proof for the WAM With Types}.
\newblock In E.~{B\"{o}rger}, G.~{J\"{a}ger}, H.~{Kleine B\"{u}ning}, and M.~M.
  Richter, editors, {\em Computer Science Logic}, volume 626 of {\em LNCS},
  pages 15--34. Springer, 1992.
\newblock \medskip \newline The {B\"{o}rger}-Rosenzweig specification and
  correctness proof for compiling Prolog to WAM \cite {BoeRos95} is extended in
  modular fashion to the type-constraint logic programming language Protos-L
  which extends Prolog with polymorphic order-sorted (dynamic) types. In this
  paper, the notion of types and dynamic type constraints are kept abstract (as
  constraint) in order to permit applications to different constraint
  formalisms like Prolog III or CLP(R). The theorem is proved that for every
  appropriate type-constraint logic programming system L, every compiler to the
  WAM extension with an abstract notion of types which satisfies the specified
  conditions, is correct. \cite{BeiBoe96b} extends the specification and the
  correctness proof to the full Protos Abstract Machine by refining the
  abstract type constraints to the polymorphic order-sorted types of PROTOS-L.
  Also issued as IBM Germany Science Center Research Report IWBS 205 , 1991.
  Revised and final version published in \cite{BeiBoe96a}.

\bibitem{BeiBoe96b}
C.~Beierle and E.~B{\"{o}}rger.
\newblock {Refinement of a typed {WAM} extension by polymorphic order-sorted
  types}.
\newblock {\em Formal Aspects of Computing}, 8(5):539--564, 1996.
\newblock \medskip \newline Continuation of \cite{BeiBoe96a} which is extended
  to the full Protos Abstract Machine by refining the abstract type constraints
  to the polymorphic order-sorted types of PROTOS-L. Preliminary version
  published under the title {\em A WAM Extension for Type-Constraint Logic
  Programming: Specification and Correctness Proof} as Research Report IWBS
  200, IBM Germany Science Center, Heidelberg, December 1991.

\bibitem{BeiBoe96a}
C.~Beierle and E.~B{\"{o}}rger.
\newblock {Specification and correctness proof of a {WAM} extension with
  abstract type constraints}.
\newblock {\em Formal Aspects of Computing}, 8(4):428--462, 1996.
\newblock \medskip \newline Revised version of \cite{BeiBoe92}.

\bibitem{BeBoDuGR96}
C.~Beierle, E.~B\"orger, I.~Durdanovic, U.~Gl\"asser, and E.~Riccobene.
\newblock {Refining Abstract Machine Specifications of the Steam Boiler Control
  to Well Documented Executable Code}.
\newblock In J.-R. Abrial, E.~B\"orger, and H.~Langmaack, editors, {\em {Formal
  Methods for Industrial Applications. Specifying and Programming the
  Steam-Boiler Control}}, number 1165 in LNCS, pages 62--78. Springer, 1996.
\newblock \medskip \newline The steam-boiler control specification problem is
  used to illustrate how ASMs applied to the specification and the verification
  of complex systems can be exploited for a reliable and well documented
  development of executable, but formally inspectable and systematically
  modifiable code. A hierarchy of stepwise refined abstract machine models is
  developed, the ground version of which can be checked for whether it
  faithfully reflects the informally given problem. The sequence of machine
  models yields various abstract views of the system, making the various design
  decisions transparent, and leads to a $C^{++}$ program. This program has been
  demonstrated during the Dagstuhl-Meeting on Methods for Semantics and
  Specification, in June 1995, to control the FZI Steam-Boiler simulator
  satisfactorily. \\ The proofs of properties of the ASM models provide insight
  into the structure of the system which supports easily maintainable
  extensions and modifications of both the abstract specification and the
  implementation. For a continuation of this line of research see
  \cite{BoeMea97}.

\bibitem{BelRic97}
G.~Bella and E.~Riccobene.
\newblock {Formal Analysis of the Kerberos Authentication System}.
\newblock {\em Journal of Universal Computer Science}, 3(12), 1997.
\newblock \medskip \newline A formal model of the whole system is reached
  through stepwise refinements of ASMs, and is used as a basis both to discover
  the minimum assumptions to guarantee the correctness of the system, and to
  analyse its security weaknesses. Each refined model comes together with a
  correctness refinement theorem.

\bibitem{Blakley92}
B.~Blakley.
\newblock {\em {A Smalltalk Evolving Algebra and its Uses}}.
\newblock PhD thesis, University of Michigan, Ann Arbor, Michigan, 1992.
\newblock \medskip \newline An early student work on ASMs (the late date of
  1992 is accidental). A reduced version of Smalltalk is formalized and
  studied.

\bibitem{BlaGur97}
A.~Blass and Y.~Gurevich.
\newblock {The Linear Time Hierarchy Theorems for Abstract State Machines}.
\newblock {\em Journal of Universal Computer Science}, 3(4):247--278, 1997.
\newblock \medskip \newline Contrary to polynomial time, linear time badly
  depends on the computation model. In 1992, Neil Jones designed a couple of
  computation models where the linear-speed-up theorem fails and linear-time
  computable functions form a proper hierarchy. However, the linear time of
  Jones' models is too restrictive. Linear-time hierarchy theorems for random
  access machines and ASMs are proven. In particular it is shown that there
  exists a sequential ASM {\em U} (an allusion to ``universal'') and a constant
  {\em c} such that, under honest time counting, {\em U} simulates every other
  sequential ASM in lock-step with log factor {\em c}. The generalization for
  ASMs is harder and more important because of the greater flexibility of the
  ASM model. One long-term goal of this line or research is to prove linear
  lower bounds for linear time problems. The result has been anounced unter the
  title {\em Evolving Algebras and Linear Time Hierarchy} in B. Pehrson and I.
  Simon (Eds.), IFIP 13th World Computer Congress, vol.I:
  Technology/Foundations, Elsevier, Amsterdam, 1994, 383-390.

\bibitem{BlGuSh97}
A.~Blass, Y.~Gurevich, and S.~Shelah.
\newblock {Choiceless Polynomial Time}.
\newblock Technical Report CSE-TR-338-97, EECS Dept., University of Michigan,
  1997.
\newblock \medskip \newline The question "Is there a computation model whose
  machines do not distinguish between isomorphic structures and compute exactly
  polynomial time properties?" became a central question of finite model
  theory. The negative answer was conjectured in \cite{Gurevich88a}. A related
  question is what portion of Ptime can be naturally captured by a computation
  model (when inputs are arbitrary finite structures). A Ptime version of ASMs
  is used to capture the portion of Ptime where algorithms are not allowed
  arbitrary choice but parallelism is allowed and, in some cases, implements
  choice.

\bibitem{Boerger90a}
E.~B{\"o}rger.
\newblock {A Logical Operational Semantics for Full Prolog. Part I: Selection
  Core and Control}.
\newblock In E.~{B\"{o}rger}, H.~{Kleine B\"{u}ning}, M.~M. Richter, and
  W.~{Sch\"{o}nfeld}, editors, {\em CSL'89. 3rd Workshop on Computer Science
  Logic}, volume 440 of {\em LNCS}, pages 36--64. Springer, 1990.
\newblock \medskip \newline See Comments to \cite{Boerger92}.

\bibitem{Boerger90b}
E.~B{\"o}rger.
\newblock {A Logical Operational Semantics of Full Prolog. Part II: Built-in
  Predicates for Database Manipulation}.
\newblock In B.~Rovan, editor, {\em Mathematical Foundations of Computer
  Science}, volume 452 of {\em LNCS}, pages 1--14. Springer, 1990.
\newblock \medskip \newline See Comments to \cite{Boerger92}.

\bibitem{Boerger92}
E.~B{\"o}rger.
\newblock {A Logical Operational Semantics for Full Prolog. Part III: Built-in
  Predicates for Files, Terms, Arithmetic and Input-Output}.
\newblock In Y.~Moschovakis, editor, {\em Logic From Computer Science},
  volume~21 of {\em Berkeley Mathematical Sciences Research Institute
  Publications}, pages 17--50. Springer, 1992.
\newblock \medskip \newline This paper, along with \cite{Boerger90a} and
  \cite{Boerger90b} are the original 3 papers of {B\"{o}rger} where he gives a
  complete ASM formalization of Prolog with all features discussed in the
  international Prolog standardization working group (WG17 of ISO/IEC JTCI
  SC22), see \cite{BoeDas90}. The specification is developed by stepwise
  refinement, describing orthogonal language features by modular rule sets. An
  improved (tree instead of stack based) version is found in
  \cite{BoeRos91c,BoeRos94}; the revised final version is in \cite{BoeRos94}.
  These three papers were also published in 1990 as IBM Germany Science Center
  Research Reports 111, 115 and 117 respectively. The refinement technique is
  further developed in \cite{BoeRos95,BoeDur96,BoeMaz97,BoeMea97,BoeSch98a}.

\bibitem{Boerger94}
E.~B{\"o}rger.
\newblock {Logic Programming: The Evolving Algebra Approach.}
\newblock In B.~Pehrson and I.~Simon, editors, {\em IFIP 13th World Computer
  Congress}, volume I: Technology/Foundations, pages 391--395, Elsevier,
  Amsterdam, the Netherlands, 1994.
\newblock \medskip \newline Surveys the work which has been done from
  1986--1994 on specifications of logic programming systems by ASMs.

\bibitem{Boerger95a}
E.~B{\"o}rger.
\newblock {Annotated Bibliography on Evolving Algebras}.
\newblock In E.~B{\"o}rger, editor, {\em Specification and Validation Methods},
  pages 37--51. Oxford University Press, 1995.
\newblock \medskip \newline An annotated bibliography of papers (as of 1994)
  which deal with or use ASMs.

\bibitem{Boerger95b}
E.~B{\"o}rger.
\newblock {Why Use Evolving Algebras for Hardware and Software Engineering?}
\newblock In M.~Bartosek, J.~Staudek, and J.~Wiederman, editors, {\em
  Proceedings of SOFSEM'95, 22nd Seminar on Current Trends in Theory and
  Practice of Informatics}, volume 1012 of {\em LNCS}, pages 236--271.
  Springer, 1995.
\newblock \medskip \newline A presentation of the salient features of ASMs, as
  part of a discussion and survey of the use of ASMs in design and analysis of
  hardware and software systems. The leading example is elaborated in detail in
  \cite{BoeMaz97}.

\bibitem{Boerger96}
E.~B\"orger.
\newblock {Evolving Algebras and Parnas Tables}.
\newblock In H.~Ehrig, F.~von Henke, J.~Meseguer, and M.~Wirsing, editors, {\em
  {Specification and Semantics}}. Dagstuhl Seminar No. 9626, July 1996.
\newblock \medskip \newline Extended abstract showing that Parnas' approach to
  use function tables for precise program documentation can be generalized and
  gentilized in a natural way by using ASMs for well-documented program
  development.

\bibitem{BoeDas90}
E.~B{\"o}rger and K.~D{\"a}ssler.
\newblock {Prolog: DIN Papers for Discussion}.
\newblock ISO/IEC JTCI SC22 WG17 Prolog Standardization Document~58, National
  Physical Laboratory, Middlesex, England, 1990.
\newblock \medskip \newline A version of \cite{Boerger90a,Boerger90b,Boerger92}
  proposed to the International Prolog Standardization Committee as a complete
  formal semantics of Prolog. An improved version is in \cite{BoeRos94}.

\bibitem{BoeDel95}
E.~B{\"o}rger and G.~{Del Castillo}.
\newblock {A formal method for provably correct composition of a real-life
  processor out of basic components ({T}he {APE100} {R}everse {E}ngineering
  {S}tudy)}.
\newblock In B.~Werner, editor, {\em Proceedings of the First IEEE
  International Conference on Engineering of Complex Computer Systems
  (ICECCS'95)}, pages 145--148, November 1995.
\newblock \medskip \newline Presents a technique, based on ASMs, by which a
  behavioural description of a processor is obtained as result of the
  composition of its (formally specified) basic architectural components. The
  technique is illustrated on the example of a subset the zCPU processor (used
  as control unit of the APE100 parallel architecture). A more complete
  version, containing the full formal description of the zCPU components, of
  their composition and of the whole zCPU processor, appeared in Y.~Gurevich
  and E.~B{\"o}rger (Eds.), {\em Evolving Algebras -- Mini-Course, BRICS
  Technical Report (BRICS-NS-95-4)}, 195-222, University of Aarhus, Denmark,
  July 1995.

\bibitem{BoDeGR94}
E.~B{\"o}rger, G.~{Del Castillo}, P.~Glavan, and D.~Rosenzweig.
\newblock {Towards a Mathematical Specification of the APE100 Architecture: the
  APESE Model.}
\newblock In B.~Pehrson and I.~Simon, editors, {\em IFIP 13th World Computer
  Congress}, volume I: Technology/Foundations, pages 396--401, Elsevier,
  Amsterdam, the Netherlands, 1994.
\newblock \medskip \newline Defines an ASM model of the high-level programmer's
  view of the APE100 parallel architecture. This simple model is refined in
  \cite{BoeDel95} to an ASM processor model.

\bibitem{BoeDem91}
E.~B{\"o}rger and B.~Demoen.
\newblock {A Framework to Specify Database Update Views for Prolog}.
\newblock In M.~J. Maluszynski, editor, {\em PLILP'91. Third International
  Symposium on Programming Languages Implementation and Logic Programming.},
  volume 528 of {\em LNCS}, pages 147--158. Springer, 1991.
\newblock \medskip \newline Provides a precise definition of the major Prolog
  database update views (immediate, logical, minimal, maximal), within a
  framework closely related to \cite{Boerger90a,Boerger90b,Boerger92}. A
  preliminary version of this was published as {\em The View on Database
  Updates in Standard Prolog: A Proposal and a Rationale} in ISO/ETC JTCI SC22
  WG17 Prolog Standardization Report no. 74, February 1991, pp 3-10.

\bibitem{BoeDur96}
E.~B{\"o}rger and I.~Durdanovi\'c.
\newblock {Correctness of compiling Occam to Transputer code}.
\newblock {\em Computer Journal}, 39(1):52--92, 1996.
\newblock \medskip \newline The final draft version has been issued in BRICS
  Technical Report (BRICS-NS-95-4), see \cite{BoeGur95}. Sharpens the
  refinement method of \cite{BoeRos95} to cope also with parallelism and non
  determinism for an imperative programming language. The paper provides a
  mathematical definition of the Transputer Instruction Set architecture for
  executing Occam together with a correctness proof for a general compilation
  schema of Occam programs into Transputer code. \smallskip \\ Starting from
  the Occam model developed in \cite{BoDuRo94}, constituted by an abstract
  processor running a high and a low priority queue of Occam processes (which
  formalizes the semantics of Occam at the abstraction level of atomic Occam
  instructions), increasingly more refined levels of Transputer semantics are
  developed, proving correctness (and when possible also completeness) for each
  refinement step. \smallskip \\ Along the way proof assumptions are collected,
  thus obtaining a set of natural conditions for compiler correctness, so that
  the proof is applicable to a large class of compilers. The formalization of
  the Transputer instruction set architecture has been the starting point for
  applications of the ASM refinement method to the modeling of other
  architectures (see \cite{BoeDel95,BoeMaz97}).

\bibitem{BoDuRo94}
E.~B{\"o}rger, I.~Durdanovi{\'c}, and D.~Rosenzweig.
\newblock {Occam: Specification and Compiler Correctness. Part I: Simple
  Mathematical Interpreters.}
\newblock In U.~Montanari and E.~R. Olderog, editors, {\em Proc. PROCOMET'94
  (IFIP Working Conference on Programming Concepts, Methods and Calculi)},
  pages 489--508. North-Holland, 1994.
\newblock \medskip \newline A truly concurrent ASM model of Occam is defined as
  basis for a provably correct, smooth transition to the Transputer Instruction
  Set architecture. This model is stepwise refined, in a provably correct way,
  providing: (a) an asynchronous implementation of synchronous channel
  communication, (b) its optimization for internal channels, (c) the sequential
  implementation of processors using priority and time--slicing. See
  \cite{BoeDur96} for the extension of this work to cover the compilation to
  Transputer code.

\bibitem{BoeGla94}
E.~B{\"o}rger and U.~Gl{\"a}sser.
\newblock {A Formal Specification of the PVM Architecture}.
\newblock In B.~Pehrson and I.~Simon, editors, {\em IFIP 13th World Computer
  Congress}, volume I: Technology/Foundations, pages 402--409, Elsevier,
  Amsterdam, the Netherlands, 1994.
\newblock \medskip \newline Provides an ASM model for the Parallel Virtual
  machine (PVM, the Oak Ridge National Laboratory software system that serves
  as a general purpose environment for heterogeneous distributed computing).
  The model defines PVM at the C--interface, at the level of abstraction which
  is tailored to the programmer's understanding. Cf. the survey {\em An
  abstract model of the parallel virtual machine (PVM)} presented at {\em 7th
  International Conference on Parallel and Distributed Computing Systems}
  (PDCS'94), Las Vegas/Nevada, 5.-9.10.1994. See \cite{BoeGla95} for an
  elaboration of this paper.

\bibitem{BoeGla95}
E.~B{\"o}rger and U.~Gl{\"a}sser.
\newblock {M}odelling and {A}nalysis of {D}istributed and {R}eactive {S}ystems
  using {E}volving {A}lgebras.
\newblock In Y.~Gurevich and E.~B{\"o}rger, editors, {\em Evolving Algebras --
  Mini-Course, BRICS Technical Report (BRICS-NS-95-4)}, pages 128--153.
  University of Aarhus, Denmark, July 1995.
\newblock \medskip \newline This is a tutorial introduction into the ASM
  approach to design and verification of complex computing systems. The salient
  features of the methodology are explained by showing how one can develop from
  scratch an easily understandable and transparent ASM model for PVM, the
  widespread virtual architecture for heterogeneous distributed computing.

\bibitem{BoGlMu94}
E.~B{\"o}rger, U.~Gl{\"a}sser, and W.~M{\"u}ller.
\newblock {The Semantics of Behavioral VHDL'93 Descriptions}.
\newblock In {\em EURO-DAC'94. European Design Automation Conference with
  EURO-VHDL'94}, pages 500--505, Los Alamitos, California, 1994. IEEE CS Press.
\newblock \medskip \newline Provides a transparent but precise ASM definition
  of the signal behavior and time model of full {\em elaborated} VHDL'93. This
  includes guarded signals, delta and time delays, the two main propagation
  delay modes {\em transport,\/inertial}, and the three process suspensions
  (wait on/until/for). Shared variables, postponed processes and rejection
  pulse are covered. The work is extended in \cite{BoGlMu95}.

\bibitem{BoGlMu95}
E.~B{\"o}rger, U.~Gl{\"a}sser, and W.~M{\"u}ller.
\newblock {Formal Definition of an Abstract VHDL'93 Simulator by EA-Machines}.
\newblock In C.~{Delgado Kloos} and P.~T. Breuer, editors, {\em Formal
  Semantics for VHDL}, pages 107--139. Kluwer Academic Publishers, 1995.
\newblock \medskip \newline Extends the work in \cite{BoGlMu94} by including
  the treatment of variable assignments and of value propagation by ports. This
  ASM model for VHDL is extended to analog VHDL in \cite{SaMiSa97}.

\bibitem{BoeGur95}
E.~B{\"o}rger and Y.~Gurevich.
\newblock {Evolving Algebras -- Mini Course}.
\newblock In E.~B{\"o}rger and Y.~Gurevich, editors, {\em BRICS Technical
  Report (BRICS-NS-95-4)}, pages 195--222. University of Aarhus, 1995.
\newblock \medskip \newline Contains reprints of the papers
  \cite{BlaGur97,Gurevich91,Gurevich94c,Gurevich94b,GurHug94,
  GurHug97,GurHug93,BoGuRo94,BoeDel95,BoeDur96,BoeGla95}.

\bibitem{BoGuRo94}
E.~B{\"o}rger, Y.~Gurevich, and D.~Rosenzweig.
\newblock {The Bakery Algorithm: Yet Another Specification and Verification}.
\newblock In E.~B{\"o}rger, editor, {\em Specification and Validation Methods},
  pages 231--243. Oxford University Press, 1995.
\newblock \medskip \newline One ASM A1 is constructed to reflect faithfully the
  algorithm. Then a more abstract ASM A2 is constructed. It is checked that A2
  is safe and fair, and that A1 correctly implements A2. The proofs work for
  atomic as well as, mutatis mutandis, for durative actions.

\bibitem{BoLoRo94}
E.~B{\"o}rger, F.~J. {L{\'o}pez-Fraguas}, and M.~{Rodr{\'i}guez-Artalejo}.
\newblock {A Model for Mathematical Analysis of Functional Logic Programs and
  their Implementations}.
\newblock In B.~Pehrson and I.~Simon, editors, {\em IFIP 13th World Computer
  Congress}, volume I: Technology/Foundations, pages 410--415, 1994.
\newblock \medskip \newline Defines an ASM model for the innermost version of
  the functional logic programming language BABEL, extending the Prolog model
  of \cite{BoeRos94} by rules which describe the reduction of expressions to
  normal form. The model is stepwise refined towards a mathematical
  specification of the implementation of Babel by a graph--narrowing machine.
  The refinements are proved to be correct. A full version containing
  optimizations and proofs appeared under the title {\em Towards a Mathematical
  Specification of a Narrowing Machine} as research report DIA 94/5, Dpto.
  Inform{\'a}tica y Autom{\'a}tica, Universidad Complutense, Madrid 1994.

\bibitem{BoeMaz97}
E.~B{\"o}rger and S.~Mazzanti.
\newblock {A Practical Method for Rigorously Controllable Hardware Design}.
\newblock In J.P. Bowen, M.B. Hinchey, and D.~Till, editors, {\em {ZUM'97: The
  Z Formal Specification Notation}}, volume 1212 of {\em LNCS}, pages 151--187.
  Springer, 1996.
\newblock \medskip \newline A technique for specifying and verifying the
  control of pipelined microprocessors is described, illustrated through formal
  models for Hennessy and Patterson's RISC architecture DLX. A sequential DLX
  model is stepwise refined to the pipelined DLX which is proved to be correct.
  Each refinement deals with a different pipelining problem (structural
  hazards, data hazards, control hazards) and the methods for its solution.
  This makes the approach applicable to design-driven verification as well as
  to the verification-driven design of RISC machines. A preliminary version
  appeared under the title {\em A correctness proof for pipelining in RISC
  architectures} as DIMACS (Rutgers University, Princeton University, ATT Bell
  Laboratories, Bellcore) research report TR 96-22, pp.1-60, Brunswick, New
  Jersey, 1995.

\bibitem{BoeMea97}
E.~B\"orger and L.~Mearelli.
\newblock {Integrating ASMs into the Software Development Life Cycle}.
\newblock {\em Journal of Universal Computer Science}, 3(5):603--665, 1997.
\newblock \medskip \newline Presents a structured software engineering method
  which allows the software engineer to control efficiently the {\em modular
  development} and the {\em maintenance} of well documented, formally
  inspectable and smoothly modifiable code out of rigorous ASM {\em models for
  requirement specifications}. Shows that the code properties of interest (like
  correctness, safety, liveness and performance conditions) can be proved at
  high levels of abstraction by traditional and reusable mathematical arguments
  which---where needed---can be computer verified. Shows also that the proposed
  method is appropriate for dealing in a rigorous but transparent manner with
  hardware-software co-design aspects of system development.\\ The approach is
  illustrated by developing a $C^{++}$ program for the production cell case
  study. The program has been validated by extensive experimentation with the
  FZI production cell simulator in Karlsruhe and has been submitted for
  inspection to the Dagstuhl seminar on ``Practical Methods for Code
  Documentation and Inspection'' (May 1997). Source code (the ultimate
  refinement) for the case study appears in \cite{Mearelli97}; the model
  checking results for the ASM models appears in \cite{Winter97}.

\bibitem{BoeRic91}
E.~B{\"o}rger and E.~Riccobene.
\newblock {Logical Operational Semantics of Parlog. Part I: And-Parallelism}.
\newblock In H.~Boley and M.~M. Richter, editors, {\em Processing Declarative
  Knowledge}, volume 567 of {\em Lecture Notes in Artificial Intelligence},
  pages 191--198. Springer, 1991.
\newblock \medskip \newline See comment to \cite{BoeRic93}.

\bibitem{BoeRic92b}
E.~B{\"o}rger and E.~Riccobene.
\newblock {A Mathematical Model of Concurrent Prolog}.
\newblock Research Report CSTR-92-15, Dept. of Computer Science, University of
  Bristol, Bristol, England, 1992.
\newblock \medskip \newline An ASM formalization of Ehud Shapiro's Concurrent
  Prolog. Adaptation of the model defined for PARLOG in \cite{BoeRic93}.

\bibitem{BoeRic92a}
E.~B{\"o}rger and E.~Riccobene.
\newblock {Logical Operational Semantics of Parlog. Part II: Or-Parallelism}.
\newblock In A.~Voronkov, editor, {\em Logic Programming}, volume 592 of {\em
  Lecture Notes in Artificial Intelligence}, pages 27--34. Springer, 1992.
\newblock \medskip \newline See comment to \cite{BoeRic93}.

\bibitem{BoeRic93}
E.~B{\"o}rger and E.~Riccobene.
\newblock {A Formal Specification of Parlog}.
\newblock In M.~Droste and Y.~Gurevich, editors, {\em Semantics of Programming
  Languages and Model Theory}, pages 1--42. Gordon and Breach, 1993.
\newblock \medskip \newline An ASM formalization of Parlog, a well known
  parallel version of Prolog. This formalization separates explicitly the two
  kinds of parallelism occurring in Parlog: AND--parallelism and
  OR--parallelism. It uses an implementation independent, abstract notion of
  terms and substitutions. Improved and extended version of
  \cite{BoeRic91,BoeRic92a}, obtained combining the concurrent features of the
  Occam model of \cite{GurMos90} with the logic programming model of
  \cite{BoeRos94}. Also published as Technical Report TR 1/93 from Dipartmento
  di Informatica, Universit{\`a} da Pisa, 1993.

\bibitem{BoeRic94}
E.~B{\"o}rger and E.~Riccobene.
\newblock {Logic + Control Revisited: An Abstract Interpreter for G{\"o}del
  Programs}.
\newblock In G.~Levi, editor, {\em Advances in Logic Programming Theory}.
  Oxford University Press, 1994.
\newblock \medskip \newline Develops a simple ASM interpreter for G{\"o}del
  programs. This interpreter abstracts from the deterministic and sequential
  execution strategies of Prolog \cite{BoeRos95} and thus provides a precise
  interface between logic and control components for execution of G{\"o}del
  programs. The construction is given in abstract terms which cover the general
  logic programming paradigm and allow for concurrency.

\bibitem{BoeRos91c}
E.~B{\"o}rger and D.~Rosenzweig.
\newblock {A Formal Specification of Prolog by Tree Algebras}.
\newblock In V.~\u{C}eric, V.~Dobri{\'c}, V.~Lu\u{z}ar, and R.~Paul, editors,
  {\em Information Technology Interfaces}, pages 513--518. University Computing
  Center, Zagreb, Zagreb, 1991.
\newblock \medskip \newline Prompted by discussion in the international Prolog
  standardization committee (ISO/IEC JTC1 SC22 WG17), this paper suggests to
  replace the stack based model of \cite{Boerger90a} and the stack
  implementation of the tree based model of \cite{Boerger90b} by a pure tree
  model for Prolog. An improved version of the latter is the basis for
  \cite{BoeRos94} where also an error in the treatment of the {\em catch}
  built-in predicate is corrected.

\bibitem{BoeRos91a}
E.~B{\"o}rger and D.~Rosenzweig.
\newblock {An Analysis of Prolog Database Views and their Uniform
  Implementation}.
\newblock Research Report CSE-TR-89-91, EECS Dept., University of Michigan, Ann
  Arbor, Michigan, 1991.
\newblock \medskip \newline A mathematical analysis of the Prolog database
  views defined in \cite{BoeDem91}. The analysis is derived by stepwise
  refinement of the stack model for Prolog from \cite{BoeRos95}. It leads to
  the proposal of a uniform implementation of the different views which
  discloses the tradeoffs between semantic clarity and efficiency of database
  update view implementations. Also issued by the international Prolog
  Standardization Committee as ISO/IEC JTCI SC22 WG17 document no. 80, National
  Physical Laboratory, Teddington, England 1991.

\bibitem{BoeRos91b}
E.~B{\"o}rger and D.~Rosenzweig.
\newblock {From Prolog Algebras Towards WAM -- A Mathematical Study of
  Implementation}.
\newblock In E.~B{\"o}rger, H.~{Kleine B{\"u}ning}, M.~M. Richter, and
  W.~Sch{\"o}nfeld, editors, {\em CSL'90, 4th Workshop on Computer Science
  Logic}, volume 533 of {\em LNCS}, pages 31--66. Springer, 1991.
\newblock \medskip \newline Refines B{\"o}rger's Prolog model \cite{Boerger90b}
  by elaborating the conjunctive component---as reflected by compilation of
  clause structure into WAM code---and the disjunctive component---as reflected
  by compilation of predicate structure into WAM code. The correctness proofs
  for these refinements include last call optimization, determinacy detection
  and virtual copying of dynamic code. Extended in \cite{BoeRos92} and improved
  in \cite{BoeRos95}.

\bibitem{BoeRos92}
E.~B{\"o}rger and D.~Rosenzweig.
\newblock {WAM Algebras -- A Mathematical Study of Implementation, Part 2}.
\newblock In A.~Voronkov, editor, {\em Logic Programming}, volume 592 of {\em
  Lecture Notes in Artificial Intelligence}, pages 35--54. Springer, 1992.
\newblock \medskip \newline Refines the Prolog model of \cite{BoeRos91b} by
  elaborating the WAM code for representation and unification of terms. The
  correctness proof for this refinement includes environment trimming, Warren's
  variable classification and switching instructions. Improved in
  \cite{BoeRos95}. Also issued as Technical Report CSE-TR-88-91 from EECS Dept,
  University of Michigan, Ann Arbor, Michigan, 1991.

\bibitem{BoeRos93}
E.~B{\"o}rger and D.~Rosenzweig.
\newblock {The Mathematics of Set Predicates in Prolog}.
\newblock In G.~Gottlob, A.~Leitsch, and D.~Mundici, editors, {\em
  Computational Logic and Proof Theory}, volume 713 of {\em LNCS}, pages 1--13.
  Springer, 1993.
\newblock \medskip \newline Provides a logical (proof--theoretical)
  specification of the solution collecting predicates {\em findall, bagof\/} of
  Prolog. This abstract definition allows a logico--mathematical analysis,
  rationale and criticism of various proposals made for implementations of
  these predicates (in particular of {\em setof\/}) in current Prolog systems.
  Foundational companion to section 5, on solution collecting predicates, in
  \cite{BoeRos94}. Also issued as {\em Prolog. Copenhagen papers 2}, ISO/IEC
  JTC1 SC22 WG17 Standardization report no. 105, National Physical Laboratory,
  Middlesex, 1993, pp. 33-42.

\bibitem{BoeRos94}
E.~B{\"o}rger and D.~Rosenzweig.
\newblock {A Mathematical Definition of Full Prolog}.
\newblock In {\em Science of Computer Programming}, volume~24, pages 249--286.
  North-Holland, 1994.
\newblock \medskip \newline An abstract ASM specification of the semantics of
  Prolog, rigorously defining the international ISO 1995 Prolog standard by
  stepwise refinement. Revised and final version of
  \cite{Boerger90a,Boerger90b,BoeDas90,BoeRos91c}. An abstract of this was
  issued as {\em Full Prolog in a Nutshell} in {\em Logic Programming\/}
  (Proceedings of the 10th International Conference on Logic Programming) (D.
  S. Warren, Ed.), MIT Press 1993. A preliminary version appeared under the
  title {\em A Simple Mathematical Model for Full Prolog} as research report
  TR-33/92, Dipartimento di Informatica, Universit{\`a} di Pisa, 1992.

\bibitem{BoeRos95}
E.~B{\"o}rger and D.~Rosenzweig.
\newblock {The WAM -- Definition and Compiler Correctness}.
\newblock In C.~Beierle and L.~Pl{\"u}mer, editors, {\em Logic Programming:
  Formal Methods and Practical Applications}, Studies in Computer Science and
  Artificial Intelligence, chapter~2, pages 20--90. North-Holland, 1994.
\newblock \medskip \newline Substantial example of the successive refinement
  method in the area, improving \cite{Boerger90a,Boerger90b,Boerger92} and the
  direct predecessors \cite{BoeRos91b,BoeRos92}. A hierarchy of ASMs provides a
  solid foundation for constructing provably correct compilers from Prolog to
  WAM. Various refinement steps take care of different distinctive features
  (``orthogonal components'' in the authors' metaphor) of WAM making the
  specification as well as the correctness proof modular and extendible;
  examples of such extensions are found in
  \cite{BeiBoe96b,BeiBoe96a,BoeSal94,Araujo97,Kwon97}. An extension of this
  work to an imperative language with parallelism and non determinism has been
  provided in \cite{BoeDur96}. See \cite{Ahrendt95,Pusch96,SchAhr97} for
  machine checked versions of the correctness proofs (for some of) the
  refinement steps. A preliminary version appeared as Research Report TR-14/92,
  Dipartimento di Informatica, Universit{\`a} di Pisa, 1992.

\bibitem{BoeSal94}
E.~B{\"o}rger and R.~Salamone.
\newblock {CLAM Specification for Provably Correct Compilation of CLP(${\cal
  R}$) Programs}.
\newblock In E.~B{\"o}rger, editor, {\em Specification and Validation Methods},
  pages 97--130. Oxford University Press, 1995.
\newblock \medskip \newline Extends the B{\"o}rger--Rosenzweig's specification
  and correctness proof, for compiling Prolog programs to the WAM
  \cite{BoeRos95}, to CLP(${\cal R}$) and the constraint logical arithmetical
  machine (CLAM) developed at IBM Yorktown Heights. For full proofs, see
  R.~Salamone, ``Una Specifica Astratta e Modulare della {CLAM} (An Abstract
  and Modular Specification of the CLAM)'', Master's Thesis, Universit{\`a} di
  Pisa, Italy, 1993.

\bibitem{BoeSch91}
E.~B{\"o}rger and P.~Schmitt.
\newblock {A Formal Operational Semantics for Languages of Type Prolog III}.
\newblock In E.~B{\"o}rger, H.~{Kleine B{\"u}ning}, M.~M. Richter, and
  W.~Sch{\"o}nfeld, editors, {\em CSL'90, 4th Workshop on Computer Science
  Logic}, volume 533 of {\em LNCS}, pages 67--79. Springer, 1991.
\newblock \medskip \newline An ASM formalization of Alain Colmerauer's
  constraint logic programming language Prolog III, obtained from the Prolog
  model in \cite{Boerger90a,Boerger90b,Boerger92} through extending
  unifications by constraint systems. This extension was the starting point for
  the extension of \cite{BoeRos95} in \cite{BeiBoe92}. A preliminary version of
  this was issued as IBM Germany IWBS Report 144, 1990.

\bibitem{BoeSch97}
E.~B\"{o}rger and P.~Schmitt.
\newblock {A Description of the Tableau Method Using Abstract State Machines}.
\newblock {\em J. Logic and Computation}, 7(5):661--683, 1997.
\newblock \medskip \newline Starting from the textbook formulation of the
  tableau calculus, the authors give an operational description of the tableau
  method in terms of ASMs at various levels of refinement ending after four
  stages at a specification that is very close to the {\mbox{{\sf
  lean}$T^{\!\!\textstyle A}\!\!P$}}~implementation of the tableau calculus in
  Prolog. Proofs of correctness and completeness of the refinement steps are
  given.

\bibitem{BoeSch98b}
E.~B\"orger and W.~Schulte.
\newblock {A Modular Design for the Java VM architecture}.
\newblock In E.~B\"orger, editor, {\em {Architecture Design and Validation
  Methods}}. Springer, 1998.
\newblock \medskip \newline Provides a modular definition of the Java VM
  architecture, at different layers of abstraction. The layers partly reflect
  the layers made explicit in the specification of the Java language in
  \cite{BoeSch98a}. The ASM model for JVM defined here and the ASM model for
  Java defined in \cite{BoeSch98a} provide a rigorous framework for a machine
  independent mathematical analysis of the language and of its implementation,
  including compilation correctness conditions, safety and optimization issues.

\bibitem{BoeSch98a}
E.~B\"orger and W.~Schulte.
\newblock {Programmer Friendly Modular Definition of the Semantics of Java}.
\newblock In J.~Alves-Foss, editor, {\em {Formal Syntax and Semantics of
  Java}}, LNCS. Springer, 1998.
\newblock \medskip \newline Provides a system and machine independent
  definition of the semantics of the full programming language Java as it is
  seen by the Java programmer. The definition is modular, coming as a series of
  refined ASMs, dealing in succession with Java's imperative core, its object
  oriented features, exceptions and threads. The definition is intended as
  basis for the standardization of the semantics of the Java language and of
  its implementation on the Java Virtual Machine, see the ASM model for the
  Java VM in \cite{BoeSch98b}. An extended abstract has been presented to the
  IFIP WG 2.2 (University of Graz, 22.-26.9.1997) by E.B\"orger and under the
  title {\em Modular Dynamic Semantics of Java} to the Workshop on Programming
  Languages (Ahrensdorp, FEHMARN island, September 25, 1997) by W. Schulte, see
  University of Kiel, Dept. of CS Research Report Series, TR {\em Arbeitstagung
  Programmiersprachen} 1997.

\bibitem{DeDuGl96}
G.~{Del Castillo}, I.~Durdanovi\'c, and U.~Gl\"asser.
\newblock {An Evolving Algebra Abstract Machine}.
\newblock In H.~Kleine B{\"u}ning, editor, {\em {\em Proceedings of the Annual
  Conference of the European Association for Computer Science Logic
  (CSL'95)}\/}, volume 1092 of {\em LNCS}, pages 191--214. Springer, 1996.
\newblock \medskip \newline Introduces the concept of an abstract machine
  ({\small EAM\/}) as a platform for the systematic development of ASM tools
  and gives a formal definition of the {\small EAM\/} ground model in terms of
  a universal ASM. A preliminary version appeared under the title {\em
  Specification and Design of the EAM (EAM - Evolving Algebra Abstract
  Machine)} as Technical Teport tr-rsfb-96-003, Paderborn University, 1996.

\bibitem{DeDoGu97}
S.~Dexter, P.~Doyle, and Y.~Gurevich.
\newblock {Gurevich Abstract State Machines and Sch\"onhage Storage
  Modification Machines}.
\newblock {\em Journal of Universal Computer Science}, 3(4):279--303, 1997.
\newblock \medskip \newline A demonstration that, in a strong sense,
  Schoenhage's storage modification machines are equivalent to unary basic ASMs
  without external functions. The unary restriction can be removed if the
  storage modification machines are equipped with a pairing function in an
  appropriate way.

\bibitem{Diehl97}
S.~Diehl.
\newblock {Transformations of Evolving Algebras}.
\newblock In {\em {Proceedings of LIRA'97 (VIII International Conference on
  Logic and Computer Science)}}, pages 43--50, Novi Sad, Yugoslavia, September
  1997.
\newblock \medskip \newline First, constant propagation is defined as a
  transformation on ASMs. Then ASMs are extended by macro definitions and
  folding and unfolding transformations for macros are defined. Next a simple
  transformation to flatten transition rules is introduced. Finally a pass
  separation transformation for ASMs is presented. For all transformations the
  operational equivalence of the resulting ASMs with the original ASMs is
  proven. In the case of pass separation, it is shown that the results of the
  computations in the original and the transformed ASMs are equal. Next pass
  separation is applied to a simple interpreter. Finally a comparison to other
  work is given. A preliminary version appeared in 1995 as Technical Report
  02/95 of Universit{\"a}t des Saarlandes.

\bibitem{Diesen95}
D.~Diesen.
\newblock {\em {Specifying Algorithms Using Evolving Algebra. Implementation of
  Functional Programming Languages}}.
\newblock Dr. scient. degree thesis, Dept. of Informatics, University of Oslo,
  Norway, March 1995.
\newblock \medskip \newline A description of a functional interpreter for ASMs,
  with applications for functional programming languages, along with proposed
  extension to the language of ASMs.

\bibitem{FoAbYe97}
B.~Fordham, S.~Abiteboul, and Y.~Yesha.
\newblock {Evolving Databases: An Application to Electronic Commerce}.
\newblock In {\em {Proceedings of the Interational Database Engineering and
  Applications Symposium (IDEAS)}}, August 1997.
\newblock \medskip \newline The authors present a rich and extensible database
  model called "evolving databases" (EDB), with a rich and precise semantics
  based on ASMs. The authors apply EDBs to electronic commerce applications.

\bibitem{Gaieb97}
M.~Gaieb.
\newblock {G{\'e}neration de sp{\'e}cifications Centaur {\'a} partir de
  specifications Montages}.
\newblock Master's thesis, Universit\'e de Nice -- Sophia Antipolis, June 1997.
\newblock \medskip \newline This works investigate the possibilities of mapping
  the operational ASM semantics of the static analysis phase of Montages
  \cite{KutPie97a} into the declarative Natural Semantics framework. A
  formalization for the list arrows of Montages is found --- a feature that has
  not been fully formalized in \cite{KutPie97a}. In addition, the Gem-Mex
  Montages tool is interfaced to the Centaur system (which executes Natural
  Semantics specificaions), and the tool suport of Centaur is exploited in
  order to generate structural editors for languages defined with Montages.

\bibitem{Gaul95}
T.~Gaul.
\newblock {An Abstract State Machine specification of the DEC-Alpha Processor
  Family}.
\newblock Verifix Working Paper [Verifix/UKA/4], University of Karlsruhe, 1995.
\newblock \medskip \newline An ASM for the DEC-Alpha processor family, derived
  directly from the original manufacturer's handbook. The specification omits
  certain less-used instructions and VAX compatibility parts.

\bibitem{Glaesser96}
U.~Gl\"asser.
\newblock {Systems Level Specification and Modelling of Reactive Systems:
  Concepts, Methods, and Tools}.
\newblock In R.~Moreno~Diaz F.~Pichler and R.~Albrecht, editors, {\em Computer
  Aided Systems Theory--EUROCAST'95: {\em Proc.~of the Fifth International
  Workshop on Computer Aided Systems Theory (Innsbruck, Austria, May 1995})},
  volume 1030 of {\em LNCS}, pages 375--385. Springer, 1996.
\newblock \medskip \newline The paper investigates the derivation of formal
  requirements and design specifications at systems level as part of a
  comprehensive design concept for complex reactive systems. In this context
  the meaning of correctness with respect to the embedding of mathematical
  models into the physical world is discussed.

\bibitem{Glaesser97}
U.~Gl{\"a}sser.
\newblock {Combining Abstract State Machines with Predicate Transition Nets}.
\newblock In F.~Pichler and R.~Moreno-D\'{\i}az, editors, {\em {Computer Aided
  Systems Theory--EUROCAST'97 (Proc.~of the 6th International Workshop on
  Computer Aided Systems Theory, Las Palmas de Gran Canaria, Spain,
  Feb.~1997)}}, volume 1333 of {\em {LNCS}}, pages 108--122. Springer, 1997.
\newblock \medskip \newline The work investigates the formal relation between
  ASMs and Pr/TPredicate Transition (Pr/T-) Nets with the aim to integrate both
  approaches into a common framework for modeling concurrent and reactive
  system behavior, where Pr/T-nets are considered as a graphical interface for
  distributed ASMs. For the class of {\em strict Pr/T-nets} (which constitutes
  the basic form of Pr/T-nets) a transformation to distributed ASMs is given.

\bibitem{GlaKar97}
U.~Gl\"asser and R.~Karges.
\newblock {Abstract State Machine Semantics of SDL}.
\newblock {\em Journal of Universal Computer Science}, 3(12), 1997.
\newblock \medskip \newline A formal semantic model of Basic SDL-92 --
  according to the {\em ITU-T Recommendation Z.100} -- is defined in terms of
  an abstract SDL machine based on the concept of a {\em multi-agent real-time
  ASM}. The resulting interpretation model is not only mathematically precise
  but also reflects the common understanding of SDL in a direct and intuitive
  manner; it provides a {\em concise} and {\em understandable} representation
  of the complete dynamic semantics of Basic SDL-92. Moreover, the model can
  easily be {\em extended} and{\em modified}. The article considers the
  behavior of channels, processes and timers with respect to signal transfer
  operations and timer operations.

\bibitem{GlaRos93}
P.~Glavan and D.~Rosenzweig.
\newblock {Communicating Evolving Algebras}.
\newblock In E.~B{\"o}rger, H.~{Kleine B{\"u}ning}, G.~J{\"a}ger, S.~Martini,
  and M.~M. Richter, editors, {\em Computer Science Logic}, volume 702 of {\em
  Lecture Notes in Computer Science}, pages 182--215. Springer, 1993.
\newblock \medskip \newline A theory of concurrent computation within the
  framework of ASMs is developed, generalizing
  \cite{GurMos90,BoeRic91,BoeRic92a,BoeRic92b}. As illustration models are
  given for the Chemical Abstract Machine and the $\pi$-calculus. See
  \cite{Gurevich94b} for a more satisfactory definition of the notion of
  distributed ASM runs.

\bibitem{GlaRos94}
P.~Glavan and D.~Rosenzweig.
\newblock {Evolving Algebra Model of Programming Language Semantics}.
\newblock In B.~Pehrson and I.~Simon, editors, {\em IFIP 13th World Computer
  Congress}, volume I: Technology/Foundations, pages 416--422, Elsevier,
  Amsterdam, the Netherlands, 1994.
\newblock \medskip \newline Defines an ASM interpretation of many-step SOS,
  denotational semantics and Hoare logic for the language of while--programs
  and states correctness and completeness theorems, based on a simple flowchart
  model of the language.

\bibitem{GoKaSc91}
G.~Gottlob, G.~Kappel, and M.~Schrefl.
\newblock {Semantics of Object-Oriented Data Models -- The Evolving Algebra
  Approach}.
\newblock In J.~W. Schmidt and A.~A. Stogny, editors, {\em Next Generation
  Information Technology}, volume 504 of {\em LNCS}, pages 144--160. Springer,
  1991.
\newblock \medskip \newline Uses ASMs to define the operational semantics of
  object creation, of overriding and dynamic binding, and of inheritance at the
  type level (type specialization) and at the instance level (object
  specialization).

\bibitem{GraGur97}
E.~Gr\"adel and Y.~Gurevich.
\newblock {Metafinite Model Theory}.
\newblock In {\em {Logic and Computational Complexity, Selected Papers}},
  number 960 in LNCS, pages 313--366. Springer, 1995.
\newblock \medskip \newline A closer look reveals that computer systems, {\em
  e.g.} databases, are not necessarily finite because they may involve for
  example arithmetic. Motivated by such computer science challenges and by ASM
  applications, metafinite structures are defined and the approach and methods
  of finite model theory are extended to metafinite models. The relevance to
  the ASM methodology:\enspace ASM states are metafinite structures. An early
  version has been presented under the title {\em Towards a Model Theory of
  Metafinite Structures} to the Logic Colloquium 1994, see the abstract in the
  {\em Journal of Symbolic Logic}. A revised version is going to appear in 1998
  in a special issue of {\em Information and Computation}.

\bibitem{GroRen95}
R.~Groenboom and G.~{Renardel de Lavalette}.
\newblock {A Formalization of Evolving Algebras}.
\newblock In {\em Proceedings of Accolade95}. Dutch Research School in Logic,
  1995.
\newblock \medskip \newline The authors present the syntax and semantics for a
  Formal Language for Evolving Algebra (FLEA). This language is then extended
  to a multi-modal language FLEA' and it is sketched how we can transfer the
  axioms of the logic MLCM to FLEA'. MLCM is a Modal Logic of Creation and
  Modification based on QDL as presented by Harel.

\bibitem{Gurevich88b}
Y.~Gurevich.
\newblock {Algorithms in the World of Bounded Resources}.
\newblock In R.~Herken, editor, {\em {The Universal Turing Machine -- A
  Half-Century Story}}, pages 407--416. Oxford University Press, 1988.
\newblock \medskip \newline Early complexity theoretical motivation for the
  introduction of ASMs is discussed.

\bibitem{Gurevich88a}
Y.~Gurevich.
\newblock {Logic and the Challenge of Computer Science}.
\newblock In E.~B{\"o}rger, editor, {\em Current Trends in Theoretical Computer
  Science}, pages 1--57. Computer Science Press, 1988.
\newblock \medskip \newline The introduction and the first use of ASMs (at the
  end of the paper).

\bibitem{Gurevich91}
Y.~Gurevich.
\newblock {Evolving Algebras. A Tutorial Introduction}.
\newblock {\em Bulletin of EATCS}, 43:264--284, 1991.
\newblock \medskip \newline The first tutorial on ASMs. The ASM thesis is
  stated: Every algorithm can be simulated by an appropriate ASM in lock-step
  on the natural abstraction level of the algorithm. A slightly revised version
  of this was reprinted in G. Rozenberg and A. Salomaa Eds, {\em Current Trends
  in Theoretical Computer Science}, World Scientific, 1993, pp 266-292. For a
  more advanced definition see \cite{Gurevich94b}.

\bibitem{Gurevich94c}
Y.~Gurevich.
\newblock {Evolving Algebras}.
\newblock In B.~Pehrson and I.~Simon, editors, {\em IFIP 13th World Computer
  Congress}, volume I: Technology/Foundations, pages 423--427, Elsevier,
  Amsterdam, the Netherlands, 1994.
\newblock \medskip \newline The opening talk at the first ASM workshop.
  Sections: Introduction, The ASM Thesis, Remarks, Future Work.

\bibitem{Gurevich94a}
Y.~Gurevich.
\newblock {Logic Activities in Europe}.
\newblock {\em ACM SIGACT News}, 1994.
\newblock \medskip \newline A critical analysis of European logic activities in
  computer science. The part relevant to ASMs is subsection~4.6 called
  Mathematics and Pedantics.

\bibitem{Gurevich94b}
Y.~Gurevich.
\newblock {Evolving Algebras 1993: Lipari Guide}.
\newblock In E.~B{\"o}rger, editor, {\em Specification and Validation Methods},
  pages 9--36. Oxford University Press, 1995.
\newblock \medskip \newline The tutorial \cite{Gurevich91} covered basic ASMs.
  In the meantime, ASMs have been extensively used, in particular, for
  specifying parallel, distributed computations and computations involving real
  time. It became obvious that a more advanced definition of ASMs is needed.
  The guide addresses this need. For a recent update {\em May 1997 Draft of the
  ASM Guide} see the Technical Report CSE-TR-336-97, EECS Dept., University of
  Michigan.

\bibitem{GurHug93}
Y.~Gurevich and J.~Huggins.
\newblock {The Semantics of the C Programming Language}.
\newblock In E.~B{\"o}rger, H.~{Kleine B{\"u}ning}, G.~J{\"a}ger, S.~Martini,
  and M.~M. Richter, editors, {\em Computer Science Logic}, volume 702 of {\em
  LNCS}, pages 274--309. Springer, 1993.
\newblock \medskip \newline The method of successive refinements (inspired by
  its application in \cite{Boerger90a,Boerger90b}) is used to give a succint
  dynamic semantics of the C programming language. For a correction of minor
  errors and omissions see the ERRATA in LNCS 832 (1994), 334-336. An early
  version appeared under the title {\em The Evolving Algebra Semantics of C:
  Preliminary Version} as Technical Report CSE-TR-141-92, EECS Department,
  University of Michigan, Ann Arbor, 1992. This work is included in the PhD
  thesis {\em Evolving Algebras: Tools for Specification, Verification, and
  Program Transformation} of the second author, pp.IX+91, University of
  Michigan, Ann Arbor, 1995. For an extension to C++ see \cite{Wallace94}.

\bibitem{GurHug94}
Y.~Gurevich and J.~Huggins.
\newblock {Evolving Algebras and Partial Evaluation}.
\newblock In B.~Pehrson and I.~Simon, editors, {\em IFIP 13th World Computer
  Congress}, volume I: Technology/Foundations, pages 587--592, Elsevier,
  Amsterdam, the Netherlands, 1994.
\newblock \medskip \newline The paper describes an automated partial evaluator
  for sequential ASMs implemented at the University of Michigan. It takes an
  ASM and a portion of its input and produces a specialized ASM using the
  provided input to execute rules when possible and generating new rules
  otherwise. A full version appears as J.~Huggins, ``An Offline Partial
  Evaluator for Evolving Algebras'', Technical Report CSE-TR-229-95, EECS
  Department, University of Michigan, Ann Arbor, 1995. This work is included in
  the PhD thesis {\em Evolving Algebras: Tools for Specification, Verification,
  and Program Transformation} of the second author, pp.IX+91, University of
  Michigan, Ann Arbor, 1995.

\bibitem{GurHug96}
Y.~Gurevich and J.~Huggins.
\newblock {The Railroad Crossing Problem: An Experiment with Instantaneous
  Actions and Immediate Reactions}.
\newblock In {\em Proceedings of CSL'95 (Computer Science Logic)}, volume 1092
  of {\em LNCS}, pages 266--290. Springer, 1996.
\newblock \medskip \newline An ASM solution for the railroad crossing problem.
  The paper experiments with agents that perform instantaneous actions in
  continuous time and in particular with agents that fire at the moment they
  are enabled. A preliminary version appeared under the title {\em The Railroad
  Crossing Problem: An Evolving Algebra Solution} as research report LITP 95/63
  of Centre National de la Recherche Scientifique, Paris, and under the title
  {\em The Generalized Railroad Crossing Problem: An Evolving Algebra Based
  Solution} as research report CSE-TR-230-95 of EECS Department, University of
  Michigan, Ann Arbor, MI. For a relation to model checking see
  \cite{BeaSli97a,BeaSli97b}.

\bibitem{GurHug97}
Y.~Gurevich and J.~Huggins.
\newblock {Equivalence Is In The Eye Of The Beholder}.
\newblock {\em Theoretical Computer Science}, 179(1-2):353--380, 1997.
\newblock \medskip \newline A response to a paper of Leslie Lamport,
  ``Processes are in the Eye of the Beholder'' which is published in the same
  volume. It is discussed how the same two algorithms may and may not be
  considered equivalent. In addition, a direct proof is given of an appropriate
  equivalence of two particular algorithms considered by Lamport. A preliminary
  version appeared as research report CSE-TR-240-95, EECS Dept., University of
  Michigan, Ann Arbor, Michigan 1995.

\bibitem{GurMan94}
Y.~Gurevich and R.~Mani.
\newblock {Group Membership Protocol: Specification and Verification}.
\newblock In E.~B{\"o}rger, editor, {\em Specification and Validation Methods},
  pages 295--328. Oxford University Press, 1995.
\newblock \medskip \newline An interesting and useful protocol of Flavio
  Cristian involves timing constraints and its correctness is not obvious. The
  protocol is formally specified and verified. (The verification proof allowed
  the authors to simplify the assumptions slightly.).

\bibitem{GurMos90}
Y.~Gurevich and L.~Moss.
\newblock {Algebraic Operational Semantics and Occam}.
\newblock In E.~B{\"o}rger, H.~{Kleine B{\"u}ning}, and M.~M. Richter, editors,
  {\em CSL'89, 3rd Workshop on Computer Science Logic}, volume 440 of {\em
  LNCS}, pages 176--192. Springer, 1990.
\newblock \medskip \newline The first application of ASMs to distributed
  parallel computing with the challenge of true concurrency. See
  \cite{BoDuRo94,BoeDur96}.

\bibitem{GuSoWa97}
Y.~Gurevich, N.~Soparkar, and C.~Wallace.
\newblock {Formalizing Database Recovery}.
\newblock {\em Journal of Universal Computer Science}, 3(4):320--340, 1997.
\newblock \medskip \newline A database recovery algorithm (the undo-redo
  algorithm) is modeled at several levels of abstraction, with verification of
  the correctness of each model. An updated version of \cite{WaGuSo95} and of
  the Technical Reports CSE-TR-249-95 and CSE-TR-327-97 of EECS Department,
  University of Michigan, Ann Arbor.

\bibitem{GurSpi97}
Y.~Gurevich and M.~Spielmann.
\newblock {Recursive Abstract State Machines}.
\newblock {\em Journal of Universal Computer Science}, 3(4):233--246, 1997.
\newblock \medskip \newline The authors suggest a definition of recursive ASMs
  in terms of distributed ASMs. Preliminary version appeared as Technical
  Report CSE-TR-322-96, EECS Department, University of Michigan, Ann Arbor,
  1996.

\bibitem{Huggins94}
J.~Huggins.
\newblock {Kermit: Specification and Verification}.
\newblock In E.~B{\"o}rger, editor, {\em Specification and Validation Methods},
  pages 247--293. Oxford University Press, 1995.
\newblock \medskip \newline The Kermit file-transfer protocol (including a
  sliding windows extension to the basic protocol) is specified and verified
  using ASMs at several different layers of abstraction. This work is included
  in the PhD thesis {\em Evolving Algebras: Tools for Specification,
  Verification, and Program Transformation} of the second author, pp.IX+91,
  University of Michigan, Ann Arbor, 1995.

\bibitem{Huggins96}
J.~Huggins.
\newblock {Broy-Lamport Specification Problem: A Gurevich Abstract State
  Machine Solution}.
\newblock Technical Report CSE-TR-320-96, EECS Dept., University of Michigan,
  1996.
\newblock \medskip \newline An ASM solution to a specification problem
  suggested by Manfred Broy and Leslie Lamport, in conjunction with the
  Dagstuhl Workshop on Reactive Systems, held in Dagstuhl, Germany, 26-30
  September, 1994. Preliminary version appeared as Technical Report
  CSE-TR-223-94, EECS Department, University of Michigan, Ann Arbor, 1994.

\bibitem{HugVan96}
J.~Huggins and D.~{Van Campenhout}.
\newblock {Specification and Verification of Pipelining in the ARM2 RISC
  Microprocessor}.
\newblock Technical Report CSE-TR-321-96, EECS Dept., University of Michigan,
  1996.
\newblock \medskip \newline A layered ASM specification of the ARM2, one of the
  early commerical RISC microprocessors. The layered specification is used to
  prove the correctness of the ARM2's pipelining techniques. Extended abstract
  appears in {\em Proceedings of the IEEE International High Level Design
  Validation and Test Workshop (HLDTV'97)}, November 1997.

\bibitem{JohMos94}
D.~Johnson and L.~Moss.
\newblock {Grammar Formalisms Viewed As Evolving Algebras}.
\newblock {\em Linguistics and Philosophy}, 17:537--560, 1994.
\newblock \medskip \newline Distributed ASMs are used to model formalisms for
  natural language syntax. The authors start by defining an ASM model of
  context free derivations which abstracts from the parse tree descriptions
  used in \cite{GurMos90,BoeRic93} and from the dynamic tree generation
  appearing in \cite{BoeRos91c,BoeRos94}. Then the simple model of context free
  rules is extended to characterise in a uniform and natural way different
  context sensitive languages in terms of ASMs. See \cite{MosJoh95a,MosJoh95b}.

\bibitem{KapWil95}
A.~Kaplan and J.~Wileden.
\newblock {Formalization and Application of a Unifying Model for Name
  Management}.
\newblock In {\em {The Third ACM SIGSOFT Symposium on the Foundations of
  Software Engineering}}, volume 20(4) of {\em Software Engineering Notes},
  pages 161--172, October 1995.
\newblock \medskip \newline Presents a unifying model for name management,
  using ASMs as the specification language for the model. A preliminary version
  appeared in July 1995 as CMPSCI Technical Report 95-60 of Computer Science
  Department, University of Massachusetts, Amherst.

\bibitem{Kappel93}
A.~M. Kappel.
\newblock {Executable Specifications Based on Dynamic Algebras}.
\newblock In A.~Voronkov, editor, {\em Logic Programming and Automated
  Reasoning}, volume 698 of {\em Lecture Notes in Artificial Intelligence},
  pages 229--240. Springer, 1993.
\newblock \medskip \newline Defines a language for specification of ASMs and
  designs an abstract target machine (namely a Prolog program) which is
  specially tailored for executing ASM computations. A prototype of the
  compiler has been implemented in Prolog. For a full version see A.~M.~Kappel,
  ``Implementation of Dynamic Algebras with an Application to Prolog'',
  Master's Thesis, Universit{\"a}t Dortmund, Germany, 1990.

\bibitem{KutPie97a}
P.~Kutter and A.~Pierantonio.
\newblock {Montages: Specifications of Realistic Programming Languages}.
\newblock {\em Journal of Universal Computer Science}, 3(5):416--442, 1997.
\newblock \medskip \newline The authors introduce Montages, a version of ASMs
  specifically tailored for specifying the static and dynamic semantics of
  programming languages. Montages combine graphical and textual elements to
  yield specifications similar in structure, length, and complexity to those in
  common language manuals, but with a formal semantics. A preliminary version
  appeared in July 1996 under the title {\em Montages: Unified Static and
  Dynamic Semantics of Programming Languages} as Technical Report 118 of
  Universita de L'Aquila.

\bibitem{KutPie97b}
P.~Kutter and A.~Pierantonio.
\newblock {The Formal Specification of Oberon}.
\newblock {\em Journal of Universal Computer Science}, 3(5):443--503, 1997.
\newblock \medskip \newline A presentation of the syntax, static semantics, and
  dynamic semantics of Oberon, using ASMs and Montages \cite{KutPie97a}. The
  dynamic semantics previously appeared as P.~Kutter, ``Dynamic Semantics of
  the Oberon Programming Language'', TIK-Report 25, ETH-Z{\"u}rich, Feburary
  1997.

\bibitem{Kwon97}
K.~Kwon.
\newblock {A Structured Presenation of a Closure-Based Compilation Method for a
  Scoping Notion in Logic Programming}.
\newblock {\em Journal of Universal Computer Science}, 3(4):341--376, 1997.
\newblock \medskip \newline An extension to logic programming which permits
  scoping of procedure definitions is described at a high level of abstraction
  (using ASMs) and refined (in a provably-correct manner) to a lower level,
  building upon the method developed in \cite{BoeRos95}.

\bibitem{LisOsi96}
A.~Lisitsa and G.~Osipov.
\newblock {Evolving algebras and labelled deductive systems for the semantic
  network based reasoning}.
\newblock In {\em Proceedings of the Workshop on Applied Semiotics, ECAI'96},
  pages 5--12, August 1996.
\newblock \medskip \newline ASMs are used to present the high-level semantics
  for MIR, an AI semantic network system. Another formalization of MIR is given
  in terms of labeled deduction systems, and the two formalizations are
  compared.

\bibitem{May97}
W.~May.
\newblock {Specifying Complex and Structured Systems with Evolving Algebras}.
\newblock In {\em {TAPSOFT'97: Theory and Practice of Software Development, 7th
  International Joint Conference CAAP/FASE}}, number 1214 in LNCS, pages
  535--549. Springer, 1997.
\newblock \medskip \newline An approach is presented for specifying complex,
  structured systems with ASMs by means of aggregation and composition.

\bibitem{Mearelli97}
L.~Mearelli.
\newblock {Refining an ASM Specification of the Production Cell to $C^{++}$
  Code}.
\newblock {\em Journal of Universal Computer Science}, 3(5):666--688, 1997.
\newblock \medskip \newline Source code for the specification problem described
  in \cite{BoeMea97}.

\bibitem{Mohnen97}
M.~Mohnen.
\newblock {A Compiler Correctness Proof for the Static Link Technique by means
  of Evolving Algebras}.
\newblock {\em Fundamenta Informatica}, 29(3):257--303, 1997.
\newblock \medskip \newline The static link technique is a common method used
  by stack-based implementations of imperative programming languages. The
  author uses ASMs to prove the correctness of this well-known technique in a
  non-trivial subset of Pascal.

\bibitem{Morris88}
J.~Morris.
\newblock {\em {Algebraic Operational Semantics and Modula-2}}.
\newblock PhD thesis, University of Michigan, Ann Arbor, Michigan, 1988.
\newblock \medskip \newline The earliest formalization of a real-life language.
  The semantical description is parse-tree directed. In the meantime, the
  methodology has developed enabling more elegant descriptions, but one has to
  start somewhere. A PhD thesis under the supervision of Yuri Gurevich. An
  extended abstract appeared as Y.~Gurevich and J.~Morris, ``Algebraic
  Operational Semantics and Modula-2'', in E.~B{\"o}rger,
  H.~{Kleine~B{\"u}ning} and M.~M.~Richter, eds., {\em CSL'87, 1st Workshop on
  Computer Science Logic}, Springer LNCS 329, 1988, pp. 81-101.

\bibitem{MorPot90}
J.~Morris and G.~Pottinger.
\newblock {Ada-Ariel Semantics}.
\newblock Odyssey Research Associates, Manuscript, July 1990.

\bibitem{MosJoh95a}
L.~S. Moss and D.~E. Johnson.
\newblock {Dynamic Interpretations of Constraint-Based Grammar Formalisms}.
\newblock {\em Journal of Logic, Language, and Information}, 4(1):61--79, 1995.
\newblock \medskip \newline Extends the work of \cite{JohMos94} to grammar
  formalisms based on Kasper-Rounds logics.

\bibitem{MosJoh95b}
L.~S. Moss and D.~E. Johnson.
\newblock {Evolving Algebras and Mathematical Models of Language}.
\newblock In L.~Polos and M.~Masuch, editors, {\em {\em Applied Logic: How,
  What, and Why}}, volume 626 of {\em Synthese Library}, pages 143--175. Kluwer
  Academic Publishers, 1995.
\newblock \medskip \newline Extends the work of \cite{JohMos94} to several
  other grammar formalisms.

\bibitem{Muller94}
B.~M{\"u}ller.
\newblock {A Semantics for Hybrid Object--Oriented Prolog Systems}.
\newblock In B.~Pehrson and I.~Simon, editors, {\em IFIP 13th World Computer
  Congress}, volume I: Technology/Foundations, Elsevier, Amsterdam, the
  Netherlands, 1994.
\newblock \medskip \newline Extends the rules given in \cite{BoeRos94} for the
  user--defined core of Prolog to define the semantics of a hybrid
  object--oriented Prolog system. The definition covers the central
  object--oriented features of: object creation and deletion, data
  encapsulation, inheritance, messages, polymorphism and dynamic binding.

\bibitem{Poetzsch93}
A.~{Poetzsch-Heffter}.
\newblock {Interprocedural Data Flow Analysis based on Temporal
  Specifications}.
\newblock Technical Report 93-1397, Cornell University, Ithaca, New York, 1993.
\newblock \medskip \newline Investigates the specification of data flow
  problems by temporal logic formulas and proves fixpoint analyses correct.
  Temporal formulas are interpreted w.r.t.~programming language semantics given
  in the framework of ASMs.

\bibitem{Poetzsch94b}
A.~{Poetzsch-Heffter}.
\newblock {Comparing Action Semantics and Evolving Algebra based Specifications
  with respect to Applications}.
\newblock In {\em Proceedings of the First International Workshop on Action
  Semantics}, 1994.
\newblock \medskip \newline Action semantics is compared to ASM based language
  specifications. In particular, different aspects relevant to language
  documentation and programming tool development are discussed.

\bibitem{Poetzsch94c}
A.~{Poetzsch-Heffter}.
\newblock {Deriving Partial Correctness Logics From Evolving Algebras}.
\newblock In B.~Pehrson and I.~Simon, editors, {\em IFIP 13th World Computer
  Congress}, volume I: Technology/Foundations, pages 434--439, Elsevier,
  Amsterdam, the Netherlands, 1994.
\newblock \medskip \newline A proposal for deriving partial correctness logics
  from simple ASM models of programming languages. A basic axiom (schema) is
  derived from an ASM and is used to obtain more convenient logics.

\bibitem{Poetzsch94a}
A.~{Poetzsch-Heffter}.
\newblock {Developing Efficient Interpreters based on Formal Language
  Specifications}.
\newblock In P.~Fritzson, editor, {\em Compiler Construction}, volume 786 of
  {\em LNCS}, pages 233--247. Springer, 1994.
\newblock \medskip \newline Reports on extensions of the MAX system enabling
  the generation and refinement of interpreters based on formal language
  specifications. In these specifications, static semantics is defined by an
  attribution mechanism and dynamic semantics is defined by ASMs. Included in
  \cite{Poetzsch97}.

\bibitem{Poetzsch97}
A.~Poetzsch-Heffter.
\newblock { Prototyping Realistic Programming Languages Based On Formal
  Specifications }.
\newblock {\em { Acta Informatica }}, 34:737--772, 1997.
\newblock \medskip \newline A tool supporting the generation of
  language-specific software from specifications is presented. Static semantics
  is defined by an attribution technique (e.g.\ for the specification of flow
  graphs). The dynamic semantics is defined by ASMs. As an example, an
  object-oriented programming language with parallelism is specified. This work
  is partly based upon \cite{Poetzsch94a}.

\bibitem{Pusch96}
C.~Pusch.
\newblock {Verification of compiler correctness for the WAM}.
\newblock In J.Harrison J.~von Wright, J.Grundy, editor, {\em Theorem Proving
  in Higher Order Logics (TPHOLs'96)}, volume 1125 of {\em LNCS}, pages
  347--362. Springer, 1996.
\newblock \medskip \newline See comment to \cite{BoeRos95}.

\bibitem{Reichel96}
H.~Reichel.
\newblock {Unifying ADT and Evolving Algebra Specifications}.
\newblock {\em Bulletin of EATCS}, 59:112--126, 1996.

\bibitem{Riccobene92}
E.~Riccobene.
\newblock {\em {Modelli Matematici per Linguaggi Logici}}.
\newblock PhD thesis, University of Catania, 1992.
\newblock \medskip \newline In Italian. Systematic treatment of ASM models for
  G{\"o}del \cite{BoeRic94}, Parlog \cite{BoeRic93}, Concurrent Prolog
  \cite{BoeRic92b}, GHC, Pandora.

\bibitem{Rosenzweig94}
D.~Rosenzweig.
\newblock {Distributed Computations: Evolving Algebra Approach}.
\newblock In B.~Pehrson and I.~Simon, editors, {\em IFIP 13th World Computer
  Congress}, volume I: Technology/Foundations, pages 440--441, Elsevier,
  Amsterdam, the Netherlands, 1994.
\newblock \medskip \newline Remarks on some ASM models of concurrent and
  parallel computation.

\bibitem{SaMiSa97}
H.~Sasaki, K.~Mizushima, and T.~Sasaki.
\newblock {Semantic Validation of VHDL-AMS by an Abstract State Machine}.
\newblock In {\em Proceedings of BMAS'97 (IEEE/VIUF International Workshop on
  Behavioral Modeling and Simulation)}, pages 61--68, Arlington, VA, October
  20-21 1997.
\newblock \medskip \newline The paper extends the ASM model defined for VHDL in
  \cite{BoGlMu94,BoGlMu95}, to provide a rigorous definition of VHDL-AMS
  following the IEEE Language Reference Manual for the analogue extension of
  VHDL. Reflecting the analysis made in this paper on the BREAK statement,
  1076.1 WG will update the LRM draft (e-mail from Dr. Hisashi Sasaki, Mixed
  Signal Design Automation Sec., Analog and Mixed Signal LSI Design Dept.,
  TOSHIBA CORPORATION, Japan).

\bibitem{Sauer93}
J.~Sauer.
\newblock {\em {Wissensbasiertes L\"osen von Ablaufsplanungsproblemen durch
  explizite Heuristiken}}.
\newblock PhD thesis, Universit{\"a}t Oldenburg, 1993.
\newblock \medskip \newline Published in: Dissertationen zur K{\"u}nstlichen
  Intelligenz, Infix-Verlag, Dr. Ekkehardt Hundt, St. Augustin 1993, pp. 204.
  Uses ASMs to define the semantics for selection and elaboration of heuristics
  for computation of goal sets in the language HERA. See also J.~Sauer,
  ``Evolving Algebras for the Description of a Meta-Scheduling System'', in
  H.~Kleine~B{\"u}ning, ed., {\em Workshop der GI-Fachgruppe Logik in der
  Informatik}, Technical Report TR-RI-94-146, Universit{\"a}t Paderborn, 1994.

\bibitem{SchAhr97}
G.~Schellhorn and W.~Ahrendt.
\newblock {Reasoning about Abstract State Machines: The WAM Case Study}.
\newblock {\em Journal of Universal Computer Science}, 3(4):377--413, 1997.
\newblock \medskip \newline The authors apply the KIV (Karlsruhe Interactive
  Verifier) system to mechanically verify the proof of correctness of the
  Prolog to WAM transformation described in \cite{BoeRos95}.

\bibitem{Schoenegge95}
A.~Sch\"onegge.
\newblock {Extending Dynamic Logic for Reasoning about Evolving Algebras}.
\newblock Technical Report 49/95, Universit{\"a}t Karlsruhe, Fakult{\"a}t
  f{\"u}r Informatik, 1995.
\newblock \medskip \newline EDL, an extension of dynamic logic, is presented,
  which permits one to directly represent statements about ASMs. Such a logic
  lays the foundation for extending the KIV (Karlsruhe Interactive Verifier) to
  reason about ASMs directly.

\bibitem{SchKap91}
M.~Schrefl and G.~Kappel.
\newblock {Cooperation Contracts}.
\newblock In T.~J. Teorey, editor, {\em Proc. 10th International Conference on
  the Entity Relationship Approach}, pages 285--307. E/R Institute, 1991.
\newblock \medskip \newline The authors introduce the concept of cooperative
  message handling and use ASMs to give formal semantics.

\bibitem{Stroetmann97}
K.~Stroetmann.
\newblock {The Constrained Shortest Path Problem: A Case Study In Using ASMs}.
\newblock {\em Journal of Universal Computer Science}, 3(4):304--319, 1997.
\newblock \medskip \newline An abstract, nondeterministic form of the
  constrained shortest path problem is defined as an ASM and proven correct,
  then refined to the level of implementation.

\bibitem{Tonino97}
H.~Tonino.
\newblock {\em {A Theory of Many-sorted Evolving Algebras}}.
\newblock Ph.d. thesis, Delft University of Technology, 1997.
\newblock \medskip \newline Based on a two-valued many-sorted logic of partial
  functions (with a complete and sound Fitch-style axiomatization) a structural
  operational and a Hoare-style axiomatic semantics is given for many-sorted
  non-distributed deterministic ASMs. The SOS semantics is defined in two
  levels, one for the sequential and one for the parallel ASM constructs. Two
  (sound but not complete) Hoare-style descriptions are given, one for partial
  and one for total correctness.

\bibitem{TonVis96}
H.~Tonino and J.~Visser.
\newblock {Stepwise Refinement of an Anstract State Machine for WHNF-Reduction
  of $\lambda$-Terms}.
\newblock Technical Report 96-154, Faculty of Technical Mathematics and
  Informatics, Delft University of Technology, 1996.
\newblock \medskip \newline A series of ASMs for finding the weak head normal
  form (WHNF) of an arbitrary term of the $\lambda$-calculus is presented.

\bibitem{Vale93}
M.~Vale.
\newblock {The Evolving Algebra Semantics of COBOL. Part I: Programs and
  Control}.
\newblock Technical Report CSE-TR-162-93, EECS Dept., University of Michigan,
  1993.
\newblock \medskip \newline An ASM for the control constructs of COBOL. A
  description of a plan for a series of ASMs for all of COBOL is sketched (but
  not implemented).

\bibitem{Visser96}
J.~Visser.
\newblock {Evolving algebras}.
\newblock Master's thesis, Faculty of Technical Mathematics and Informatics,
  Delft University of Technology, Zuidplantsoen 4, 2628 BZ Delft, The
  Netherlands, 1996.
\newblock \medskip \newline The monad programming method is used to write a
  compiler/run-analyzer for ASMs in Gofer. Static functions can be supplied to
  the ASMs by means of Gofer functions.

\bibitem{Wallace94}
C.~Wallace.
\newblock {The Semantics of the C++ Programming Language}.
\newblock In E.~B{\"o}rger, editor, {\em Specification and Validation Methods},
  pages 131--164. Oxford University Press, 1995.
\newblock \medskip \newline The semantical description in \cite{GurHug93} is
  extended to encompass all of C++.

\bibitem{Wallace97}
C.~Wallace.
\newblock {The Semantics of the Java Programming Language: Preliminary
  Version}.
\newblock Technical Report CSE-TR-355-97, EECS Dept., University of Michigan,
  December 1997.
\newblock \medskip \newline A specification of the static and dynamic semantics
  of Java, using ASMs and Montages \cite{KutPie97a}.

\bibitem{WaGuSo95}
C.~Wallace, Y.~Gurevich, and N.~Soparkar.
\newblock {Formalizing Recovery in Transaction-Oriented Database Systems}.
\newblock In S.~Chaudhuri, A.~Deshpande, and R.~Krishnamurthy, editors, {\em
  Proceedings of the Seventh International Conference on Management of Data},
  pages 166--185, New Delhi, India, 1995. Tata McGraw-Hill.
\newblock \medskip \newline The specification and verification of the Undo/Redo
  algorithm is presented in a discussion of ASMs as a formal tool for database
  recovery. An early version of \cite{GuSoWa97}.

\bibitem{Winter97}
K.~Winter.
\newblock {Model Checking for Abstract State Machines}.
\newblock {\em Journal of Universal Computer Science}, 3(5):689--701, 1997.
\newblock \medskip \newline A framework is developed for using a model checker
  to verify ASM models. It is applied to the production cell control model
  described in \cite{BoeMea97}.

\bibitem{Zamulin97c}
A.~Zamulin.
\newblock {Algebraic Specification of Dynamic Objects}.
\newblock In {\em {Proceedings of LMO'97 (Acte du Colloque Langage et Modeles a
  Objets}}, pages 111--127, Paris, 22-24 October 1997. Edition Hermes.
\newblock \medskip \newline A model for describing the behavior of dynamic
  objects is presented, using a state-transition system with the same semantics
  as (though not explicitly identified as) ASMs.

\bibitem{Zamulin97b}
A.~Zamulin.
\newblock {Specification of an Oberon Compiler by means of a Typed Gurevich
  Machine}.
\newblock Technical Report 589.3945009.00007-01, Institute of Informatics
  Systems of the Siberian Division of the Russian Academy of Sciences,
  Novosibirsk, 1997.
\newblock \medskip \newline A Typed Gurevich Machine \cite{Zamulin97a} is used
  to define a compiler for Oberon to an algebraically-specified abstract target
  machine.

\bibitem{Zamulin97a}
A.~Zamulin.
\newblock {Typed Gurevich Machines Revisited}.
\newblock Joint CS \& IIS Bulletin, Computer Science, 1997.
\newblock \medskip \newline An approach to combining type-structured algebraic
  specifications and ASMs is proposed. A preliminary version appeared in 1996
  as preprint 36 of the Institute of Informatics Systems, Novosibirsk.

\bibitem{ZimGau97}
W.~Zimmerman and T.~Gaul.
\newblock {On the Construction of Correct Compiler Back-Ends: An ASM Approach}.
\newblock {\em Journal of Universal Computer Science}, 3(5):504--567, 1997.
\newblock \medskip \newline The authors use ASMs to construct provably correct
  compiler back-ends based on realistic intermediate languages (and check the
  correctness of their proofs using PVS).

\end{thebibliography}

\end{document}